# Study of uncertainty and repeatability in structured-light 3D scanners

**María-Eugenia Polo (1\*), Aurora Cuartero (2) and Ángel M. Felicísimo (1)**

(1) Centro Universitario de Mérida, University of Extremadura, 06800 Mérida, Spain; mepolo@unex.es ; amfeli@unex.es

(2) Escuela Politécnica, University of Extremadura, 10004 Cáceres, Spain; acuartero@unex.es

(\*)Correspondence: mepolo@unex.es ; Tel.: +34-924-673-084

**Abstract**

Structured-light 3D scanners create 3D models with high accuracy, but controlling the accuracy and repeatability of the scanner is essential. The objective of this paper is to analyze the repeatability and accuracy of two structured-light 3D scanners (Go!SCAN 20TM and Go!SCAN 50TM). The method used scans steel gauge blocks several times with different resolutions to analyze the scanned data and to test the correlation between uncertainty, resolution, and scanner model. The primary results include: 1) a systematic error of magnitude similar to nominal accuracy exists and must be corrected and 2) the global uncertainty is approximately 0.05 mm without significant differences between the two scanner models. A strategy for scanning is proposed based on the results at different resolutions.

**Keywords:** accuracy; gauge block; repeatability; structured-light 3D scanner; uncertainty.

## 1. Introduction

Nowadays, diverse users are able to work with many different 3D scanners to generate models of objects or scenes with multiple purposes [1], [2], [3], [4], [5]. Most of the current 3D scanners allow the accurate, fast, and efficient creation of 3D models. Different digital scanning methods related to different technologies have led to the development of several scanning devices, such as the well-known Terrestrial Laser Scanners (TLS) [6], which are capable of operating over long distances, and the desktop scanners, which are based on the triangulation principle [7] and are a good value.

As an alternative to laser scanners, we can use highly accurate structured-light scanners. The structured-light scanner is a non-contact, optical system based on the projection of a calibrated light pattern onto the object to be scanned to capture the deformation of the pattern and generate a point cloud with the possibility of texture information [8], [9], [10].

Studying the accuracy and precision of any 3D scanner offers the opportunity to improve the scanner itself, as well as the purpose for which the 3D model was generated. Of particular importance is the growing 3D printing industry, which demands high-accuracy 3D models, and other applications that require very high-accuracy models, such as reverse engineering.

There are several kinds of work related to studying the uncertainty of 3 scanners [11], [12], [13], [14], [15] because the sources of error are varied, from instrumental biases to residual deviations of point cloud registration [16], [17]. The measure of the absolute accuracy in a 3D scanner is a difficult task because it is necessary to obtain a set of control data with very high accuracy. One option is to scan an object whose dimensions are accurately known, such as gauge blocks [18]. When it is not possible to measure the absolute accuracy, a study of repeatability can be performed [19].

The objective of this paper is to explore the repeatability and accuracy of two structured-light 3D scanners (Go!SCAN 20™ and Go!SCAN 50™). We scanned three steel gauge blocks several times with different resolutions, measuring the distances between the planes of the scanned models in each block to analyze the measurements and to test the correlation between uncertainty and resolution.

## 2. Materials

### 2.1. Scanners

The experiments presented in this paper were conducted with two portable structured-light scanners with different resolutions, Go!SCAN 20 and Go!SCAN 50, released in 2014 by the Canadian company Creaform. The distance measurement of these scanners is based on the projection of a calibrated light pattern onto the object to be scanned using white light (LED) as the light source; therefore, these devices do not have contact with the object. The scanner has a light-structured emitter and three cameras framed in a set of four LEDs; two cameras capture the deformation of the pattern and generate a point cloud, with the possibility of texture information using the third camera [Figure 1]. In this work, the texture option was disabled so the scanning process was faster, and the generated file was lighter than in a scanning session with texture.

Additional material provided by the manufacturer includes two 39 cm diameter turntables to facilitate the data collection process and adhesive circular reflecting targets. There are two types of targets, both with a black border and reflective background; the Go!SCAN 20 includes a 3 mm diameter target and the Go!SCAN 50 a 6 mm diameter target. In both cases, the scanner recognizes the targets and uses them as a positioning method to reference the 3D model. Using positioning targets is the best method to achieve accurate results [20] [21]. The manufacturer specifications of both scanners are shown in Table 1. All units are expressed in millimeters.

Table 1. Technical specifications for the scanners (Units in millimeters).

|  | Go!SCAN 20 | Go!SCAN 50 |
|---|---|---|
| Scanning area | 143 x 108 | 380 x 380 |
| Stand-off distance | 380 | 400 |
| External dimensions | 154 x 178 x 335 | 150 x 171 x 251 |
| Resolution | 0.1, 0.2, 0.3, 0.4, 0.5, 0.6, 0.7, 0.8, 0.9, 1, 2, 3, 4, 5, 6, 7, 8, 9, 10 | 0.5, 0.6, 0.7, 0.8, 0.9, 1, 2, 3, 4, 5, 6, 7, 8, 9, 10 |
| Accuracy | 0.1 | 0.1 |
| Positioning methods | Geometry, color and/or targets | |

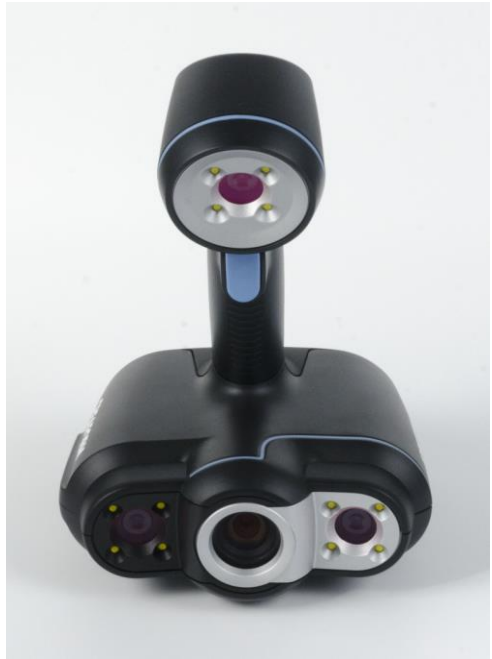

**Figure 1**. The Go!SCAN scanner has three cameras; two cameras (upper and lower right) capture the deformation of the pattern and generate a point cloud, with possibility of texture information using the third camera (left). The light-structured emitter is between the two bottom cameras.

*2.2. Gauge Blocks*

Three Mitutoyo steel, rectangular parallelepiped gauge blocks were used with nominal lengths of 100, 150 and 250 mm [Figure 2]. According to the certificate of inspection, Table 2 shows the nominal lengths and the accuracy of the blocks offered by Mitutoyo, expressed in μm.

**Table 2.** Nominal lengths and accuracy values for the blocks.

| Nominal Length (mm) | Central deviation (μm) | Maximum Deviation (μm) | Minimum Deviation (μm) | Variation (μm) |
|---|---|---|---|---|
| 100 | +0.01 | +0.03 | -0.07 | +0.10 |
| 150 | +0.20 | +0.24 | +0.12 | +0.12 |
| 250 | +0.33 | +0.36 | +0.17 | +0.19 |

The manufacturer guarantees the nominal length of the blocks at a temperature of 20 degrees Celsius. The grade of the gauge blocks (Grade 1) was selected according to the purpose of this paper, i.e., calibrating instruments.

The gauge blocks present reflective properties, so it was necessary to apply a powder of microfine talc granules suspended with a trace of lanolin using a brush pen to reduce the reflectivity.

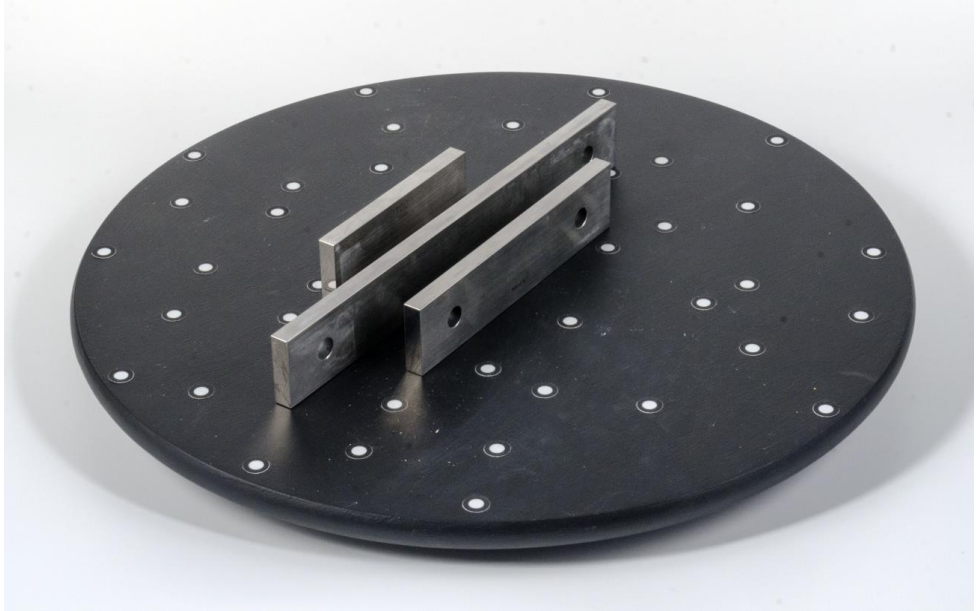

**Figure 2**. The three Mitutoyo gauge blocks with a nominal length of 100, 250, and 150 mm from left to right, respectively, on the turntable with 6 mm adhesive circular reflecting targets for the Go!SCAN50™ scanner.

*2.3. Software*

The scanners are managed by VXelements, a fully integrated 3D software platform that controls the 3D scanning process and is composed of several modules. In our case, the VXscan and VXmodel modules were used. The first module is dedicated to the acquisition and optimization of 3D scanning data. The second module allows the post-treatment creation of a mesh from the point cloud and performs a Scan-to-CAD workflow.

**3. Methods**

*3.1. Scan capture*

Measurements were made under laboratory conditions. Temperature remained constant at approximately 20 degrees Celsius throughout the scanning time. Measurements were obtained in two different sessions, one for each scanner, under dark conditions once the instrument had acclimated to the ambient temperature.

Before each scan session, it was necessary to perform a self-calibration process to allow the scanner to adapt to the environmental conditions. This process involves scanning a calibration sheet with adhesive circular reflecting targets. Each scanner has a specific self-calibration sheet with different targets, using 3 mm diameter targets for the Go!SCAN 20 and 6 mm diameter targets for the Go!SCAN 50. The scanner recognizes the targets and uses them to perform the self-calibration process at different distances.

The three gauge blocks over the turntable [Figure 2] were scanned 10 times in four different resolutions with each scanner. For Go!SCAN 20, 0.2, 0.4, 0.6 and 0.8 mm resolution were used and for Go!SCAN 50, 0.5, 0.7, 1, and 2 mm resolution were used. Therefore, 80 scans were performed in total.

*3.2. Preparing data*

As the result of any scan is a disordered cloud of points, the VXmodels module generates an aligned mesh for each scan. All further operations are carried out with these meshes.

Every gauge block is a rectangular parallelepiped. The manufacturer only assures the longitudinal dimension, i.e., 100, 150, and 250 mm. As the purpose of this paper is to check this dimension in the 3D scanned model, the normal planes to the longitudinal dimension should be defined by selecting points in the mesh and creating a fitting plane (6 planes per scan).

Here, there is a source of error because the fitting plane is obtained by performing an adjustment of the selected mesh points. The software offers the values of standard deviation and the maximum-minimum difference between points of the created fitting plane. When a generated fitting plane presented a high standard deviation value in the adjustment, the plane was removed and generated again, improving the results. This procedure was applied only a few times.

The meshes containing each plane were exported to .csv format. The .csv file contains the coordinates of the central point of the fitting plane and the unit vector perpendicular to the fitting plane.

In total, we have 480 .csv files (2 scanners x 10 replicas x 4 resolutions x 6 planes) named with four groups of digits. The first group is related to the scanner used (20 or 50), the second is the resolution expressed in tenths of millimeters, the third is the replicate number (from 01 to 10) and the last digit reports information about the block (S, Small, 100 mm; M, Medium, 150 mm, and L, Large, 250 mm) and the face of the block (A or B). The face A or B is named according to an external reference in the turntable. For example, the file, 50_02_01_MA.csv, corresponds to the first replicate of the A plane of the medium block (150 mm) scanned with 0.2 mm of resolution using Go!SCAN 50.

The next step renames the .csv files to .txt files to combine them in a unique file by replicates. This fusion was made using Text File Joiner software. The purpose is to combine the data in a spreadsheet for analysis.

The distance between each pair of fitting planes is calculated from the coordinates of the central point of the plane. Therefore, 240 distances were analyzed across the 8 datasets (2 scanners x 4 resolutions). Each dataset has 30 records (10 replicas x 3 distances).

*3.3. Data analysis*

These operations allow us to perform a data analysis related to the following points: a) errors in the adjustment of the fitting planes, b) systematic errors in the scanning process, c) analysis of uncertainty and repeatability, and d) correlation between uncertainty and resolution.

**4. Results**

*4.1. Adjustment of the fitting planes*

The fitting plane is obtained by performing an adjustment of the selected mesh points. As discussed before, this process is not free of error, but the software reports statistical values that allow evaluation and correction.

These statistical values are the standard deviation of the adjustment and the maximum-minimum difference between points for each created fitting plane (240 planes in total). The mean values, the standard deviation of the values of adjustment, and the range (R), sorted by scanner and resolutions and expressed in mm, are shown in Table 3. Resolutions are 0.2, 0.4, 0.6, and 0.8 mm for Go!SCAN 20 and 0.5, 0.7, 1.0, and 2.0 mm for Go!SCAN 50.

Table 3. Statistical values for the adjustments of the fitting planes (Units in millimeters).

| Scan 20 | 0.2 | 0.4 | 0.6 | 0.8 |
|---|---|---|---|---|
| Mean (SD) | 0.045 | 0.025 | 0.025 | 0.020 |
| Standard Dev. (SD) | 0.022 | 0.017 | 0.024 | 0.014 |
| Mean (R) | 0.697 | 0.236 | 0.194 | 0.147 |
| Standard Dev. (R) | 0.458 | 0.259 | 0.217 | 0.126 |
| Scan 50 | 0.5 | 0.7 | 1.0 | 2.0 |
| Mean (SD) | 0.039 | 0.036 | 0.037 | 0.072 |
| Standard Dev. (SD) | 0.028 | 0.016 | 0.016 | 0.036 |
| Mean (R) | 0.294 | 0.260 | 0.228 | 0.325 |
| Standard Dev. (R) | 0.232 | 0.130 | 0.098 | 0.183 |

*4.2. Systematic errors of the block measured lengths*

The differences between the true values of block lengths and the measured values show that systematic errors exist in all scans. Table 4 shows these differences between the expected value and mean value of the replicates for each block, scanner and resolution, expressed in mm.

Table 4. Systematic errors in scan (Units in millimeters and micrometers).

| Scan 20 | Small block | Medium block | Large block | Global |
|---|---|---|---|---|
| 0.2 | -0.065 | -0.054 | -0.118 | -0.079 |
| 0.4 | -0.149 | -0.085 | -0.163 | -0.132 |
| 0.6 | -0.019 | -0.042 | -0.092 | -0.051 |
| 0.8 | -0.089 | -0.065 | -0.152 | -0.102 |
| Scan 50 | Small block | Medium block | Large block | Global |
| 0.5 | +0.119 | +0.139 | +0.172 | 0.143 |
| 0.7 | +0.089 | +0.096 | +0.160 | 0.115 |
| 1.0 | +0.000 | +0.035 | +0.058 | 0.031 |
| 2.0 | -0.148 | -0.148 | -0.215 | -0.170 |

We can observe that all systematic errors from Go!SCAN 20 are negative and all errors except the last from Go!SCAN 50 are positive. None of the scanners show a recognizable relationship between systematic errors and block size values. Since the manufacturer assures 0.1 mm of accuracy in both scanners, the systematic errors should not be neglected.

*4.3. Measure uncertainty*

The uncertainty is estimated when comparing the dimension of the gauge blocks with the results of the scanned data after correcting the systematic error. Table 5 shows the standard deviation values of the errors for each block and resolution and the mean for all the blocks.

**Table 5.** Standard deviation of the errors by block and resolution (Units in millimeters and micrometers).

| Scan 20 | Small block | Medium block | Large block | Global |
|---------|-------------|--------------|-------------|--------|
| 0.2 | 0.079 | 0.051 | 0.064 | 0.065 |
| 0.4 | 0.054 | 0.042 | 0.052 | 0.049 |
| 0.6 | 0.161 | 0.025 | 0.091 | 0.092 |
| 0.8 | 0.025 | 0.021 | 0.022 | 0.023 |
| Scan 50 | Small block | Medium block | Large block | Global |
| 0.5 | 0.026 | 0.037 | 0.037 | 0.033 |
| 0.7 | 0.042 | 0.030 | 0.042 | 0.038 |
| 1.0 | 0.069 | 0.059 | 0.045 | 0.058 |
| 2.0 | 0.122 | 0.166 | 0.236 | 0.175 |

Figure 3 and Figure 4 show the values of the uncertainty (measured as the 95% confidence interval using the previous values of standard deviation) for each scanner and resolution for all the blocks. The X axis represents the resolutions in mm, and the Y axis shows the uncertainty values.

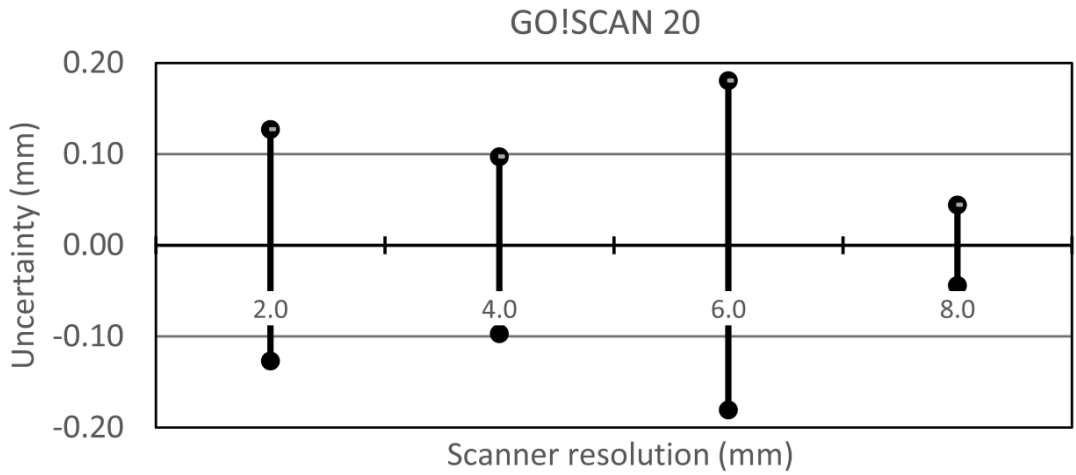

**Figure 3.** Go!SCAN 20 uncertainty versus nominal scanner resolution.

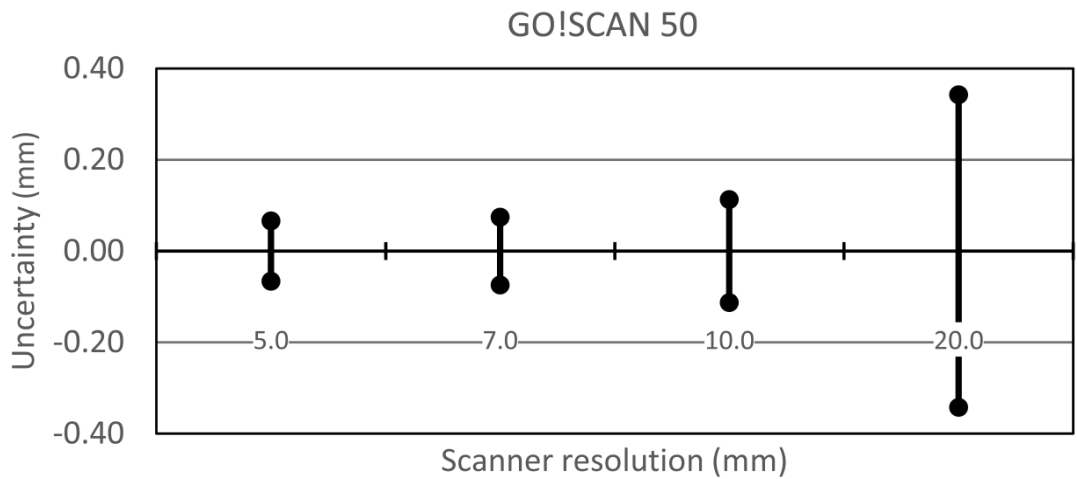

**Figure 4.** Go!SCAN 50 uncertainty versus nominal scanner resolution.

Please note that the vertical axes of both figures are different.

## 5. Conclusions

In this paper, the accuracy, repeatability, and uncertainty of two structured-light 3D scanners have been estimated. Nominal resolutions have been used and several series of measurements of gauge blocks have been performed.

The first result is that a systematic error exists in all scans. This error is negative in Go!SCAN 20 and generally positive in Go!SCAN 50. Despite this difference, it is relevant that the systematic errors have values similar to the nominal accuracy provided by the manufacturer. Obviously, it is highly recommended that analysis and correction of this error is addressed when high fidelity is required.

From the other results regarding uncertainty, it is possible to propose a strategy for scanning that is summarized as follows.

If a resolution greater than 0.5 mm is needed, use of the model 20 is mandatory.

Within the common resolution range of the two scanner models, between 0.4 and 0.8 mm, it is preferable to use the model 50. The primary reason is the similar or better uncertainty values, but we must also consider that the model 50 has a greater depth of field and is consequently more convenient and flexible. For coarse resolution values, we recommend the model 50, but with scanning using a resolution of 1 mm. If coarser values are needed, it is preferable to decimate the mesh later with a dedicated algorithm. Selecting scanner resolution values of 2 mm or greater seems inadvisable because of the increase in the uncertainty value.

**Acknowledgments:** The authors would like to thank the Government of Extremadura and European Regional Development Fund (ERDF) for supporting the publication edition and fees (GR15129).